# UNet-3D with Adaptive TverskyCE Loss for Pancreas Medical Image Segmentation

Xubei Zhang[1] and Mikhail Y. Shalaginov[2] and Tingying Helen Zeng*[3]

[1] University of Southern California, Los Angeles, CA 90089, USA
[2] Massachusetts Institute of Technology, Cambridge, MA 02139, USA
[3]*InnoBridge Institute, One Broadway, Cambridge, MA 02142, USA
*Corresponding author

**Abstract.** Pancreatic cancer, which has a low survival rate, is one of the most challenging cancers to diagnose and treat effectively. Early detection through abdominal computed tomography (CT) scans is crucial, yet complicated by the pancreas' obscure anatomical position, small size, and frequent occlusion by surrounding organs. These factors make the pancreas particularly difficult to identify and segment accurately. While deep learning (DL) models have shown promise for segmentation tasks, their performance still requires significant improvement to address these challenges. In this research, we propose a novel adaptive TverskyCE loss for DL model training, which combines Tversky loss with cross-entropy loss through learnable weights. Our method enables automatic adjustment of loss contributions during training, dynamically optimizing the objective function for improved performance. All experiments were conducted on the National Institutes of Health (NIH) Pancreas-CT dataset. We evaluated the adaptive TverskyCE loss on the UNet-3D and Dilated UNet-3D, and our method achieved a Dice Similarity Coefficient (DSC) of 85.59%, with peak performance up to 95.24%, and the $F_2$ score of 85.14%. DSC and the score $F_2$ score were improved by 9.47% and 8.98% respectively compared with the baseline UNet-3D with Tversky loss for pancreas segmentation.

**Keywords:** Pancreas segmentation · Tversky loss · Cross-entropy loss · UNet-3D · Dilated UNet-3D

## 1 Introduction

The pancreas is a vital organ with dual exocrine and endocrine functions, playing crucial roles in digestion and blood sugar regulation. Located in the upper abdomen behind the stomach and nestled between duodenum, its complex anatomical position presents significant challenges for medical imaging. Pancreatic diseases range from common conditions like diabetes and pancreatitis to lethal malignancies such as pancreatic cancer —often termed the "king of cancers" due to its aggressive nature and poor prognosis. Early-stage pancreatic tumors are particularly difficult to detect because of the small size, and by the time of diagnosis, treatment options are often severely limited [1-6]. This underscores the critical need for improved detection methods, where accurate segmentation of pancreas CT scans serves as a fundamental prerequisite for reliable diagnosis.

Convolutional Neural Networks (CNN)[7] are primarily applied to process visual data including images and graphs. UNet [8], a specialized CNN variant for medical imaging, enhances standard CNN performance for image segmentation through its symmetric expanding path and skip connections [9]. This architecture has been extended to handle volumetric data such as CT and Magnetic Resonance Imaging (MRI) scans. UNet-3D [10], changing from 2D to 3D convolutions, displays a greater performance than other models. However, UNet-based pancreas segmentation faces significant challenges due to data imbalance in training networks, where the number of pancreas voxels is much lower than non-pancreas voxels because of the organ's small size. To address this imbalance, Salehi etc.[11] introduced Tversky loss, which is a generalization function of the Dice similarity coefficient (DSC) and $F_2$ scores. This loss function achieves an improved trade-off between precision and recall through adjusting the weights of the false positives (FPs) and false negatives (FNs). For more stable optimization of UNet 3D, Giddwani et al. [12] proposed a Weighted Fusion Loss (WFL) combining Dice loss and cross-entropy (CE) loss, where the weight factor in WFL are set manually. For automatically tuning the weight factor in fusion loss according to the learning process, in this paper, we propose an adaptive TverskyCE objective function fusing the Tversky loss with CE loss, which both overcomes data-imbalance and make UNet-3D converge to a better state.

## 2 Related Work

Pancreas segmentation in medical imaging remains a challenging task due to its high anatomical variability, low contrast, and irregular shape. Recent studies have leveraged advanced deep learning techniques to improve segmentation accuracy. Roth et al. [13] proposed a probabilistic bottom-up approach using multi-level deep convolutional networks (ConvNets) to segment the pancreas in abdominal CT scans. They achieved a state-of-the-art DSC of 71.8±10.7% in testing, demonstrating the effectiveness of hierarchical ConvNets.

Another study by Milletari et al.[14] presented a fully convolutional network, V-Net, for 3D volume segmentation. This work inspired the Deep Dilated V-Net by Giddwani et al.[12], which employed a multi-rate deep dilated network in V-Net to segment the pancreas in CT images, achieving significant improvements.

Loss functions also play a crucial role in pancreas segmentation [15]. Tversky Loss, Dice Loss, and Cross-entropy Loss have been extensively studied for medical image segmentation tasks. Tversky Loss balances sensitivity and specificity by tuning its and parameters, making it suitable for imbalanced datasets. Dice Loss, a popular choice, directly optimizes the DSC, excelling in segmenting small structures. Cross-entropy Loss, though simple, works well with fully annotated data but struggles with imbalanced classes. The adaptive t-vMF Dice loss, as reported in a recent paper [16], achieved the highest average DSC compared to the original Dice loss, confirming its effectiveness in imbalanced segmentation tasks, where it automatically determines the parameter in the t-vMF similarity for adaptive t-vMF Dice loss.

Moreover, Generative Adversarial Networks (GANs) have been explored for semi-supervised learning, as shown in a study [17] that proposed a novel method to train a segmentation model with both labeled and unlabeled images. This approach prevented overfitting and improved segmentation performance, especially when few labeled images were available.

Deep learning techniques, especially UNets and their variants, along with innovative loss functions, hold significant potential in advancing pancreas segmentation research. In this paper, we propose an adaptive TverskyCE objective function that combines the Tversky loss with CE loss to overcome data-imbalance automatically and improve UNet-3D convergence. We evaluated our method on UNet-3D and Dilated UNet for 3D volume segmentation of the pancreas in CT images.

## 3 Methods

### 3.1 UNet-3D

UNet-3D, a variant of the popular UNet architecture, has emerged as a powerful tool in medical image segmentation. Specifically designed to handle three-dimensional medical image data, UNet-3D extends the original UNet's encoder-decoder structure to better capture spatial context and improve segmentation performance.

The model structure of UNet-3D mirrors its 2D predecessor, with an encoder that progressively downsamples the image to extract and compress features, and a decoder that gradually upsamples the feature maps to restore the original spatial resolution. However, unlike the 2D version, UNet-3D operates on volumetric data, such as CT or MRI scans, allowing it to capture depth information. Figure 1 shows the UNet-3D structure used in this paper for pancreas segmentation.

It consists of a contracting path (to the right) and an expanding path (to the left). To learn and use local information, high-resolution 3D features in the contracting path are concatenated with up-sampled versions of global low-resolution 3D features in the expanding path. Through this concatenation, the network learns to use both high-resolution local features and low-resolution global features. The contracting path contains padded 3×3×3 convolutions followed by ReLU non-linear layers. A 2×2×2 max pooling operation with stride 2 is applied after every two convolutional layers. After each downsampling by the max pooling layers, the number of features is doubled. In the expanding path, a 2×2×2 transposed convolution operation is applied after every two convolutional layers, and the resulting feature map is concatenated to the corresponding feature map from the contracting path. At the final layer a 1×1×1 convolution with Softmax output is used to reach the feature map with a depth equal to the number of classes (lesion or non-lesion tissue), where the loss function is calculated.

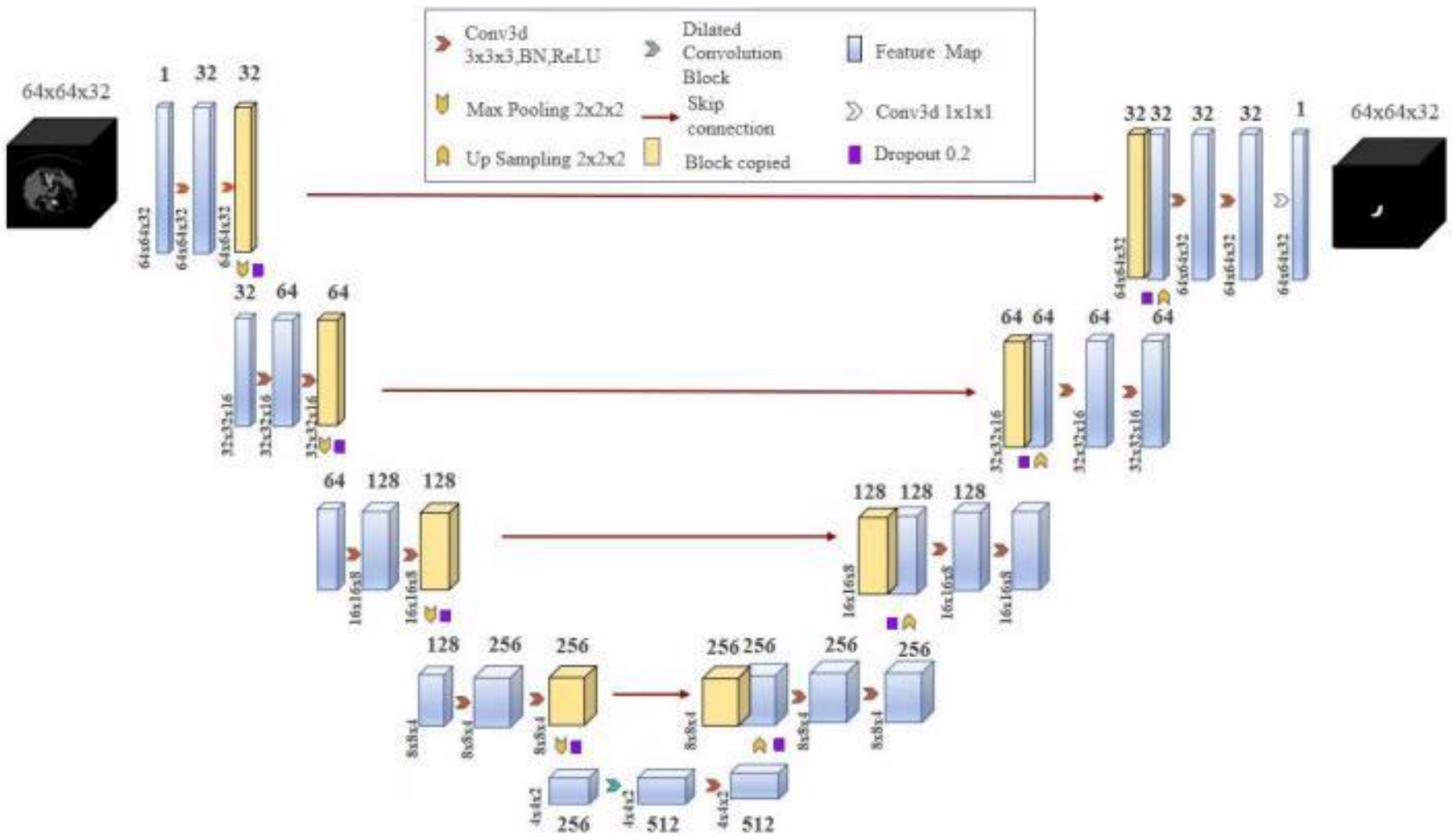

**Fig 1.** UNet-3D for pancreas segmentation

The advantages of UNet-3D include its ability to preserve spatial information through skip connections, which fuse feature maps from different layers of the encoder and decoder. These connections enable the network to leverage both low-level and high-level features for segmentation, enhancing accuracy and robustness. Additionally, UNet-3D's design facilitates the extraction of multi-scale features, critical for capturing the complex morphologies and textures often present in medical images.

One of the primary advantages of UNet-3D lies in its ability to handle 3D data natively, addressing the limitations of traditional 2D methods that lose spatial information. By operating in three dimensions, UNet-3D can better capture the spatial context of medical images, leading to improved segmentation performance. This is particularly beneficial in applications such as tumor detection, brain structure segmentation, and heart function analysis.

Moreover, UNet-3D's architecture is highly extensible and adaptable, making it suitable for a variety of medical image segmentation tasks. Its robust performance with limited training data further enhances its appeal, as medical image datasets are often small and difficult to annotate.

UNet-3D has demonstrated significant potential in medical image segmentation by leveraging its 3D capabilities and skipping connections to achieve high-quality segmentation results. Its ability to handle complex volumetric data, preserve spatial information, and adapt to various tasks makes it a valuable tool for advancing medical image analysis and diagnosis.

However, to capture different receptive fields in CNN, the work [12] proposed a network that integrated Dilated CNN with UNet-3D at the bottleneck layer. In this paper we also evaluate the performance of Dilated UNet-3D with our adaptive TverskyCE loss. Figure 2 shows the Dilated UNet-3D structure used in this paper.

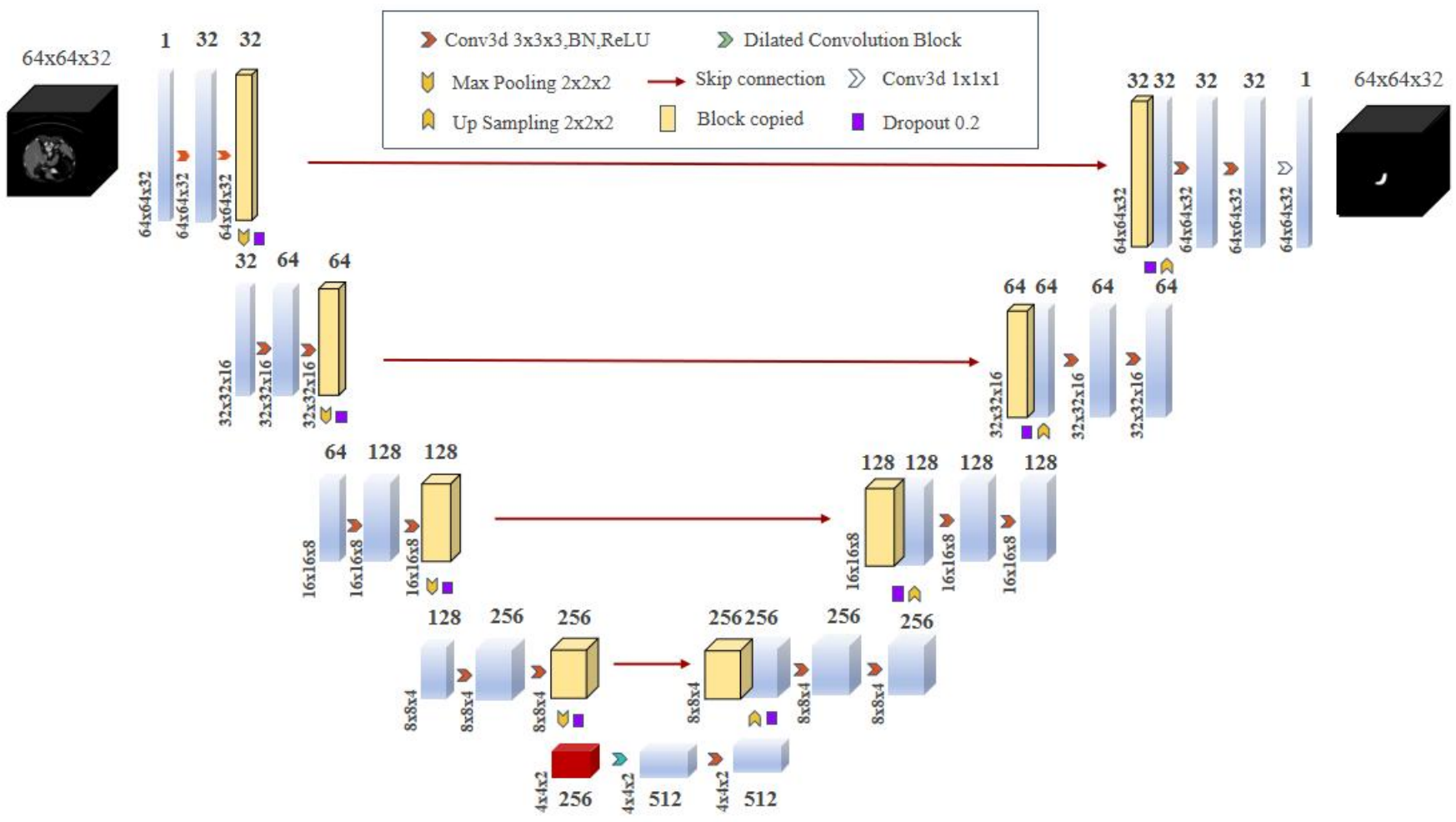

**Fig 2.** Dilated UNet-3D for pancreas segmentation

### 3.2 Adaptive TverskyCE loss

*A. Tversky loss*

After the output layer of networks, Softmax is applied across each voxel to facilitate the computation of the loss. Let P denote the set of predicted binary labels and G represent the set of ground truth binary labels. The Dice similarity coefficient D, which measures the similarity between two binary volumes, is defined as follows:

$$D(P,G) = \frac{2|PG|}{|P|+|G|} \quad (1)$$

For training networks on highly imbalanced datasets where accurately detecting small lesions is paramount, Salehi et al.[11] assigned greater weight to FNs than to FPs. The Tversky index is defined as follows:

$$S(P,G;\alpha,\beta) = \frac{|PG|}{|PG|+\alpha|P\setminus G|+\beta|G\setminus P|} \quad (2)$$

where $\alpha$ and $\beta$ control the magnitude of penalties for FPs and FNs, respectively.

The Tversky loss function is defined as the following formulation:

$$T(\alpha,\beta) = \frac{\sum_{i=1}^{N} p_{0i}g_{0i}}{\sum_{i=1}^{N} p_{0i}g_{0i} + \alpha\sum_{i=1}^{N} p_{0i}g_{1i} + \beta\sum_{i=1}^{N} p_{1i}g_{0i}} \quad (3)$$

where in the output of the Softmax layer, the $p_{0i}$ is the probability of voxel $i$ being a pancreas and $p_{1i}$ is the probability of voxel $i$ being a non-pancreas. Also, $g_{0i}$ is 1 for a pancreas voxel and 0 for a non-pancreas voxel and vice versa for the $g_{1i}$.

Employing this formulation eliminates the need to balance training weights. By adjusting the hyperparameters $\alpha$ and $\beta$, we can regulate the balance between false positives and false negatives. Notably, when $\alpha = \beta = 0.5$, the Tversky index reduces to the Dice coefficient, which is equivalent to the F1 score. With $\alpha = \beta = 1$, Equation 2 yields the Tanimoto coefficient, and setting $\alpha + \beta = 1$ generates a set of $F_\beta$ scores. Increasing $\beta$ prioritizes recall over precision by giving more weight to false negatives. We set $\beta$=0.7 in Tversky loss function during training UNet-3D to enhance generalization and performance for imbalanced datasets, effectively minimizing false negatives and boosting recall.

*B. Cross-entropy loss*

Cross-entropy generally measures the error between two probability distributions. In machine learning, it is a widely used as a loss function when the target contains class indices providing as class probabilities information [18]. The cross-entropy loss is computed between input logits and target. For a binary classification task (a classification task with two classes), a binary average cross-entropy is computed as Formula (4).

$$BCE = -\frac{1}{N}\left[\sum_{i=1}^{N}\left[t_i \log(p_i) + (1-t_i)\log(1-p_i)\right]\right] \quad (4)$$

For $N$ data samples where $t_i$ is the truth value taking a value 0 or 1 and $p_i$ is the Softmax probability for the $i^{th}$ data point. For a multi-class classification task, cross-entropy (or categorical cross-entropy as it is often referred to) can be simply extended as Formula (5).

$$CE = -\frac{1}{N}\sum_{i=1}^{N}\sum_{c=1}^{M} t_{ic}\log(p_{ic}) \quad (5)$$

where $M$ is the number of classes.

*C. Adaptive TverskyCE loss*

Although Tversky loss considers the class imbalance and involves weighted averaging of true classes, false positive classes, and false negative classes, the fluctuation of the derivative of Tversky loss is significant [19]. The gradient of Tversky loss with respect to $p_{0i}$ and $p_{1i}$ are shown as Formula (6) and (7).

$$\frac{\partial T}{\partial p_{0i}} = 2\frac{g_{0j}(\sum_{i=1}^{N} p_{0i}g_{0i} + \alpha\sum_{i=1}^{N} p_{0i}g_{1i} + \beta\sum_{i=1}^{N} p_{1i}g_{0i}) - (g_{0j} + \alpha g_{1j})\sum_{i=1}^{N} p_{0i}g_{0i}}{(\sum_{i=1}^{N} p_{0i}g_{0i} + \alpha\sum_{i=1}^{N} p_{0i}g_{1i} + \beta\sum_{i=1}^{N} p_{1i}g_{0i})^2} \quad (6)$$

$$\frac{\partial T}{\partial p_{1i}} = -\frac{\beta g_{1j}\sum_{i=1}^{N} p_{0i}g_{0i}}{(\sum_{i=1}^{N} p_{0i}g_{0i} + \alpha\sum_{i=1}^{N} p_{0i}g_{1i} + \beta\sum_{i=1}^{N} p_{1i}g_{0i})^2} \quad (7)$$

When only Tversky loss is used, the value of the loss function fluctuates greatly during training, and it is difficult to make the model converge. However, the cross-entropy loss function is continuous and differentiable, and its gradient can be readily calculated and is meaningful in most cases. This means that during the training process, the model can effectively update parameters based on the gradient of the CE loss function. Especially in machine learning, we usually tend to maximize the likelihood of the model to the data, i.e., the probability of the model predicting a given data label. The cross-entropy loss function is a special case of the negative logarithmic likelihood function under polynomial distribution (for multi classification problems) or Bernoulli distribution (for binary classification problems). Minimizing the cross-entropy loss is equivalent to maximizing the logarithmic likelihood of the data.

To improve the training efficiency and the performance of UNet-3D model, we propose an adaptive TverskyCE loss by combining Tversky loss with cross-entropy loss. The loss proportions $W$ as a factor are introduced in the total loss function. Therefore, the total loss function is as shown in Formulas (8).

$$Loss_{total}(t) = w_{Tversky}(t)L_{Tversky}(t) + w_{BCE}(t)\,L_{BCE}(t) \tag{8}$$

where

$$w_{Tversky}\,(t) = \frac{L_{Tversky}\,(t-1)}{L_{Tversky}\,(t-1)+L_{BCE}(t-1)} \tag{9}$$

$$w_{BCE}\,(t) = \frac{L_{BCE}\,(t-1)}{L_{Tversky}\,(t-1)+L_{BCE}(t-1)} \tag{10}$$

$$w_{Tversky}\,(t) + w_{BCE}\,(t) = 1 \tag{11}$$

As shown in Formulas (9) and (10), $w_{Tversky}$ and $w_{BCE}$ at the $t^{th}$ training epoch are determined by the proportion of the corresponding loss value at training epoch $t-1$ in the total loss. The loss proportion reflects the learning status of the model. The larger the proportion value of Tversky loss, the worse the ability of segmentation the model has. The larger the proportion value of BCE loss, the worse the accuracy of classification the model has. For a greater proportion in the total loss, in the next training step, the model learns automatically more about that task. Therefore, with our dynamic adjustment strategy, the UNet-3D model appropriately adjusts the learning strategy according to the loss proportion and allocates appropriate attention to each target.

# 4 Experiments

## 4.1 Datasets

All experiments are conducted on the Cancer Imaging Archive (TCIA) dataset from the National Institutes of Health (NIH) Clinical Center [20], consisting of 80 contrast-enhanced abdominal 3D CT scans. Each scan has a resolution of 512×512 pixels and a slice thickness ranging from 1.5 to 2.5 mm, with varying pixel sizes. For our work, we split the data into training, validation, and testing. The training set consists of 56 volumes of randomly selected patients, and the validation set has the other 8 volumes. The testing set consists of 16 randomly selected volumes.

### 4.2 Training

Our models are trained on 3D CT volumes of the pancreas. The implementation is done in PyTorch, using a UNet-3D as the base architecture. In the place of max-pooling layers, the convolutional kernel size is 2x2x2. An Adam optimizer with an initial learning rate of 0.005 and an adaptive decay schedule is used to facilitate stable convergence. We train each model configuration with a batch size of 10. To thoroughly investigate the effectiveness of our proposed approach, we consider six different configurations: UNet-3D (with or without dilation) using three different loss functions (Tversky, adaptive TverskyCE with α=0.3 and β=0.7 , adaptive TverskyCE with α=β=0.5). The epoch is 150.

### 4.3 Evaluation metrics

DSC and $F_2$ score were reported as primary metrics. DSC quantifies the overlap between prediction and ground truth and is widely used in medical image segmentation. The $F_2$ score, a recall-oriented metric, is particularly suitable for scenarios where minimizing false negatives is critical. Additionally, we monitor sensitivity (true positive rate), specificity (true negative rate), and precision. These metrics are defined in terms of true positives (TP), true negatives (TN), false positives (FP), and false negatives (FN) as follows:

$$Sensitivity = TP/(TP + FN) \tag{12}$$

$$Precision = TP/(TP + FP) \tag{13}$$

$$Specificity = TN/(TN + FP) \tag{14}$$

### 4.4 Experimental results

Table 1 compares the six network configurations and Dilated V-Net[12], and Table 2 provides other metrics for the six configurations. The experimental results demonstrate that the adaptive TverskyCE loss ( *α*=*β*=0.5) outperforms the traditional Tversky baseline and DiceCE. The baseline performance of the original architecture is a DSC of 76.11% and an $F_2$ score of 76.16%. Among the other five revised architectures, the lowest DSC of 72.84% and $F_2$ score of 73.69% are observed with the Dilated UNet-3D using Tversky loss. Dilated UNet-3D with adaptive TverskyCE loss achieves moderate results, with a DSC of 80.9% and $F_2$ score 80.48% for *α*=0.3 and *β*=0.7, and a DSC of 83.4% and $F_2$ score of 82.85% for *α*=*β*=0.5, respectively. Dilated UNet-3D with adaptive TverskyCE loss (*α*=*β*=0.5) has the comparable performance with the work [12]. However, the model with adapative TverskyCE automatically tunes the factors for the losses automatically. UNet-3D with adaptive TverskyCE (*α*=0.3, *β*=0.7) loss produces better performance, with a DSC of 84.18% and $F_2$ score of 83.75%. The best performance is obtained by UNet-3D with adaptive TverskyCE loss (*α*=*β*=0.5), achieving a DSC of 85.59% and a $F_2$ score of 85.14%, with the highest DSC reaching 95.2%. These results represent improvements of 9.47% and 8.98% over the baseline in terms of DSC and $F_2$ score, respectively.

**Table 1.** The Comparison on DSC and score (%)

| **Model** | Loss | $F_2$ score | DSC |
|---|---|---|---|
| **UNet-3D** | Tversky loss (*α*=0.3, *β*=0.7) | 76.16 | 76.11 |
| | adaptive TverskyCE loss (*α*=0.3, *β*=0.7) | 83.75 | 84.18 |
| | **adaptive TverskyCE loss (*α*=*β*=0.5)** | **85.14** | **85.59** |
| Dilated UNet-3D | Tversky loss (*α*=0.3, *β*=0.7) | 73.69 | 72.84 |
| | adaptive TverskyCE loss (*α*=0.3, *β*=0.7) | 80.48 | 80.9 |
| | TverskyCE loss (*α*=*β*=0.5) | 82.85 | 83.41 |
| Dilated VNet-3D [12] | TverskyCE loss (*α*=*β*=0.5) | -- | 83.3 |

**Table 2.** The Comparison on specificity (Spec.), sensitivity (Sens.) and precision (Prec.) (%)

| **Model** | Loss | Spec. | Sens. | Prec. |
|---|---|---|---|---|
| **UNet-3D** | Tversky loss (*α*=0.3, *β*=0.7) | 99.92 | 86.23 | 81.55 |
| | adaptive TverskyCE loss (*α*=0.3, *β*=0.7) | 99.98 | 84.40 | 95.73 |
| | **adaptive TverskyCE loss (*α*=*β*=0.5)** | 99.97 | 86.09 | 95.36 |
| Dilated UNet-3D | Tversky loss (*α*=0.3, *β*=0.7) | 99.74 | 88.50 | 74.91 |
| | adaptive TverskyCE loss (*α*=0.3, *β*=0.7) | 99.98 | 84 | 91.86 |
| | TverskyCE loss (*α*=*β*=0.5) | 99.98 | 84.44 | 93.64 |
| Dilated VNet-3D [12] | TverskyCE loss (*α*=*β*=0.5) | -- | 87.7 | 97 |

Notably, the adaptive approach yields a superior balance between recall (score) and precision in pancreas segmentation. The proposed UNet-3D with adaptive TverskyCE loss captures the complex boundaries of the pancreas more accurately, mitigating both under- and over-segmentation.

An example of pancreas segmentation by UNet-3D with adaptive TverskyCE loss (α= β=0.5) is shown in Fig. 3.

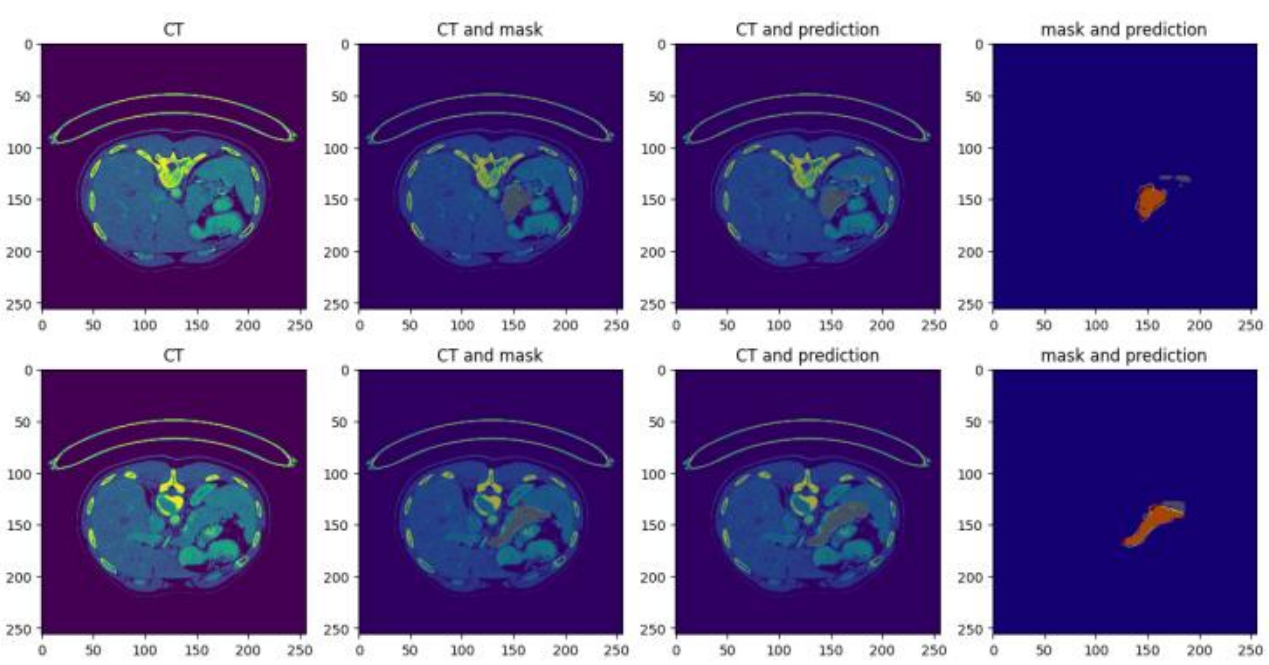

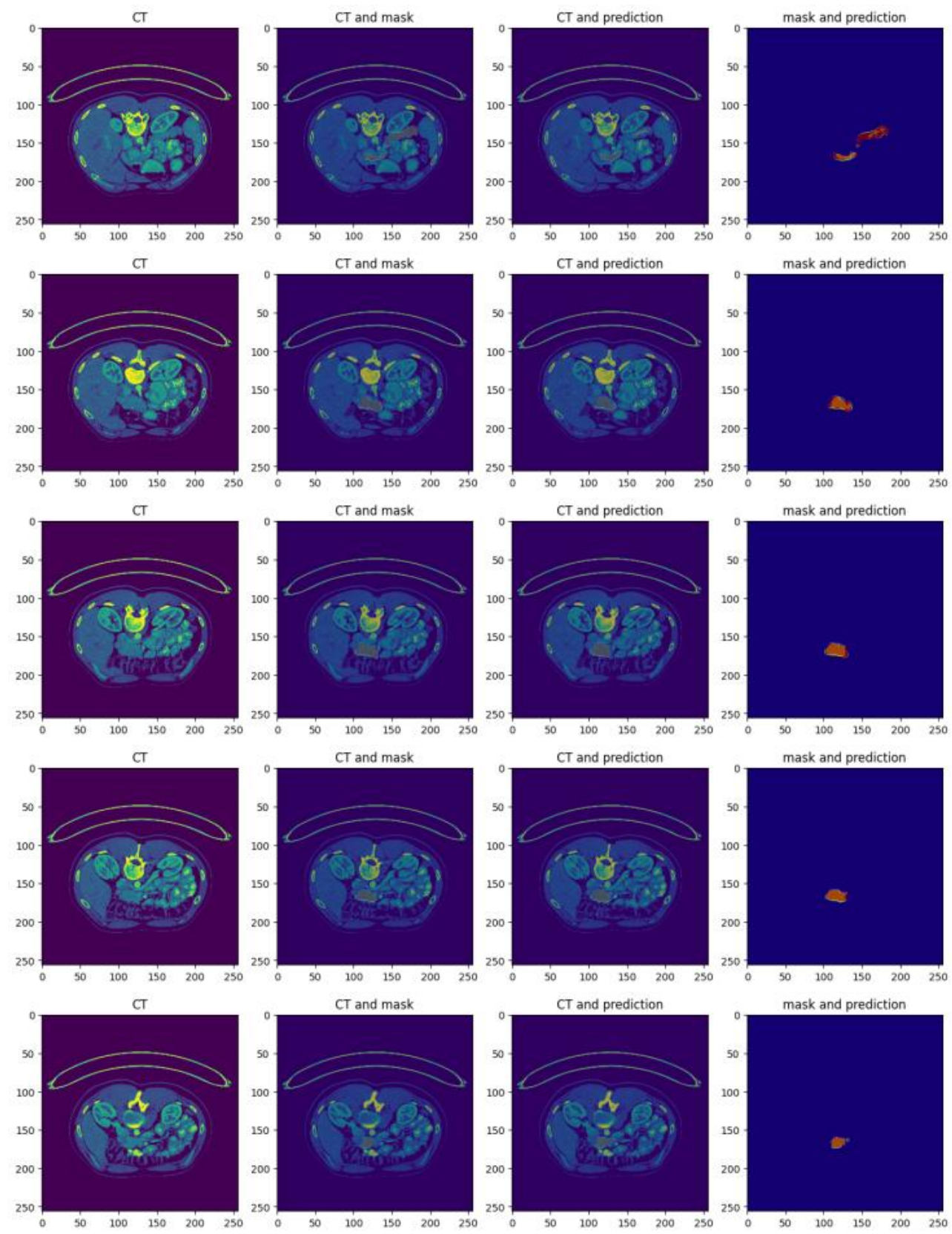

**Fig 3.** An example of pancreas segmentation by UNet-3D with adaptive TverskyCE loss ($\alpha$=$\beta$=0.5). The DSC on testing data is 95.53%.

# 5 Conclusion

In this research, we present an adaptive-loss-based UNet architecture for 3D pancreas segmentation using the TCIA datasets. By dynamically fusing Tversky and Cross-Entropy losses, our proposed adaptive TverskyCE approach effectively addresses data imbalance and enables UNet-3D to achieve higher DSC and  scores. The comparative results underscore the importance of selecting appropriate loss functions for medical image segmentation, especially when working with small and irregularly shaped structures like the pancreas. Future research includes integrating a classification step to identify pancreatic cancer and evaluating the proposed method on additional pancreas datasets. The overall improvements in segmentation accuracy and reliability highlight the significant potential of adaptive loss functions in advancing computer-aided diagnosis and guiding therapeutic decision-making.

## 6 Acknowledgements

The first author is grateful to Innobridge Institute for the Intern Research Sponsorship. (www.innobridgeinstitute.org) Thanks to Dr. Elizaveta Tremsina from MIT EECS to proof reading this paper and gave valuable inputs for the revisions.